\documentclass[epj,referee]{svjour}
\usepackage{graphics}
\begin{document}
\title{Decay Properties of the Roper Resonance from $pp \to pp\pi^+\pi^-$}
\author{H. Clement representing the PROMICE/WASA collaboration}

%
%
\institute{Physikalisches Institut der Universit\"at T\"ubingen, Morgenstelle
  14, D-72076 T\"ubingen}

\date{Received: date / Revised version: date}
%
\abstract{
Exclusive measurements of the two-pion production channel $pp \to
pp\pi^+\pi^-$ have been carried out near threshold at CELSIUS with the
PROMICE/WASA detector. They reveal $pp \to pp^\ast(1440) \to pp\sigma \to
pp(\pi^+\pi^-)_{I=\ell=0}$ as the dominant process at these energies, however,
the data exhibit also significant contributions from $p^\ast(1440) \to \Delta
\pi \to 
p(\pi^+\pi^-)_{I=\ell=0}$. From the observed interference of these 
routes their relative branching ratio is derived. 
} 
\authorrunning{H. Clement et al.}
\titlerunning{Decay Properties of the Roper Resonance from $pp \to
  pp\pi^+\pi^-$} 
\maketitle
\section{Introduction}
\label{intro}
In contrast to the $\Delta(1232)$ and other higher-lying resonances the second
excited state of the nucleon, the Roper resonance $N^\ast(1440)$ 
is still poorly understood both theoretically and experimentally
\cite{pdg00}. Being hardly observed in electromagnetic processes and  having
quantum numbers identical to those of the nucleon, 
the $N^\ast(1440)$ has been interpreted as the breathing mode monopole
excitation of the nucleon. Recent theoretical works \cite{kre00,her02} find
the Roper excitation to rest solely on meson-nucleon dynamics, whereas another 
recent investigation \cite{mor99} proposes it to be actually two resonances
with one 
being the breathing mode and the other one a $\Delta$ excitation built on top
of the $\Delta(1232)$. In all these aspects the decay modes of the Roper into
the $N\pi\pi$ channels play a crucial role. There the simplest decay is
$N^\ast \to 
N(\pi\pi)_{I=l=0} := N\sigma$, i.e., the decay into the $\sigma$ channel. A
competitive and according to present knowledge \cite{pdg00} actually much
stronger decay channel is the Roper decay into the $\Delta(1232)$ resonance
$N^\ast \to \Delta\pi$. However, this decay channel is not very well defined,
since the $\Delta$ is 
not stable and decays nearly as fast as the Roper does. In fact, most of this
decay will end up again in the $N\sigma$ channel and thus will interfere with
the direct $N^\ast \to N\sigma$ decay.

Calculations of the Valencia group \cite{alv98} predict  the $pp \to
pp\pi^+\pi^-$ reaction at energies not far above threshold  to  proceed
dominantly via 
$\sigma$ exchange in the initial $NN$ collision with successive excitation of
the Roper resonance in one of the nucleons. Our first exclusive measurements
of this reaction at $T_p = 750$ MeV \cite{bro02} support very much this
conception. This finding suggests $pp \to pp\pi\pi$
to be unique in the sense that it selectively provides the excitation
mode $``\sigma$'' N $\to N^\ast$ (where ``$\sigma$'' stands now for the
$\sigma$ 
exchange), which is not accessible in any other basic reaction process leading
to the Roper excitation.

\section{Data and Analysis}
\label{sec:2}
In Fig. 1 we present a selection of the 750 MeV data \cite{bro02} together
with new data \cite{pae02} at $T_p = 775$ MeV, again taken with the
PROMICE/WASA detector setup at CELSIUS. For experimental details see
\cite{bro02,pae02,cal96}. To see whether the reaction indeed proceeds via
$N^\ast$ excitation, we inspect the measured distribution of the invariant
mass $M_{p\pi^+\pi^-}$. At both energies the data are substantially enhanced
towards the high-energy  end compared to pure phase space (shaded areas in
Fig. 1) 
and compatible with the low-energetic tail of the $N^\ast$ excitation as
reproduced by the appropriate MC simulations. In these simulations the
amplitude for the $N^\ast$ decay is written as
$${\cal A} \sim 1 + c {\bf k}_1 \cdot {\bf k}_2 \left(3 D_{\Delta^{++}} +
D_{\Delta^0}\right) \eqno(1)$$
which in the full reaction amplitudes complements the propagators for $\sigma$
exchange and $N^\ast$ excitation as well as the expression describing the
final state interaction between the outgoing protons in relative
$s$-wave. $D_{\Delta^{++}}$ and $D_{\Delta^0}$ are the $\Delta$ propagators,
the constant 1 stands for the process $N^\ast \to N\sigma$ and the second term
for the decay route $N^\ast \to \Delta\pi \to N\sigma$, where ${\bf k}_1$ and
${\bf k}_2$ are the pion momenta. The mixing coefficent $c$ of the two decay
routes may be directly read off the data for $M_{\pi^+\pi^-}$ and
$\sigma(\delta_{\pi^+\pi^-})$, where $\delta_{\pi^+\pi^-} =~<\!\!\!)~({\bf
  k}_1,{\bf k}_2)$ is the opening angle between the two pions, see Fig. 1. The
latter distribution directly reflects the squared decay amplitude (1) averaged
over all possible pion momenta at given $\delta_{\pi^+\pi^-}$, i.e.,
$\sigma(\delta_{\pi^+\pi^-}) \sim (1 + a~\cos \delta_{\pi^+\pi^-})^2$ with $a
= c < k_1k_2 (3D_{\Delta^{++}} + D_{\Delta^0})>$, where the brackets denote
the average over all possible 
combinations. For $a \ll 1$ the distribution $\sigma(\delta_{\pi^+\pi^-})$  is
essentially linear in $a$, as exhibited by the data (Fig. 1, bottom). 

\begin{figure}
\begin{center}
\resizebox{0.2\textwidth}{!}{%
\includegraphics{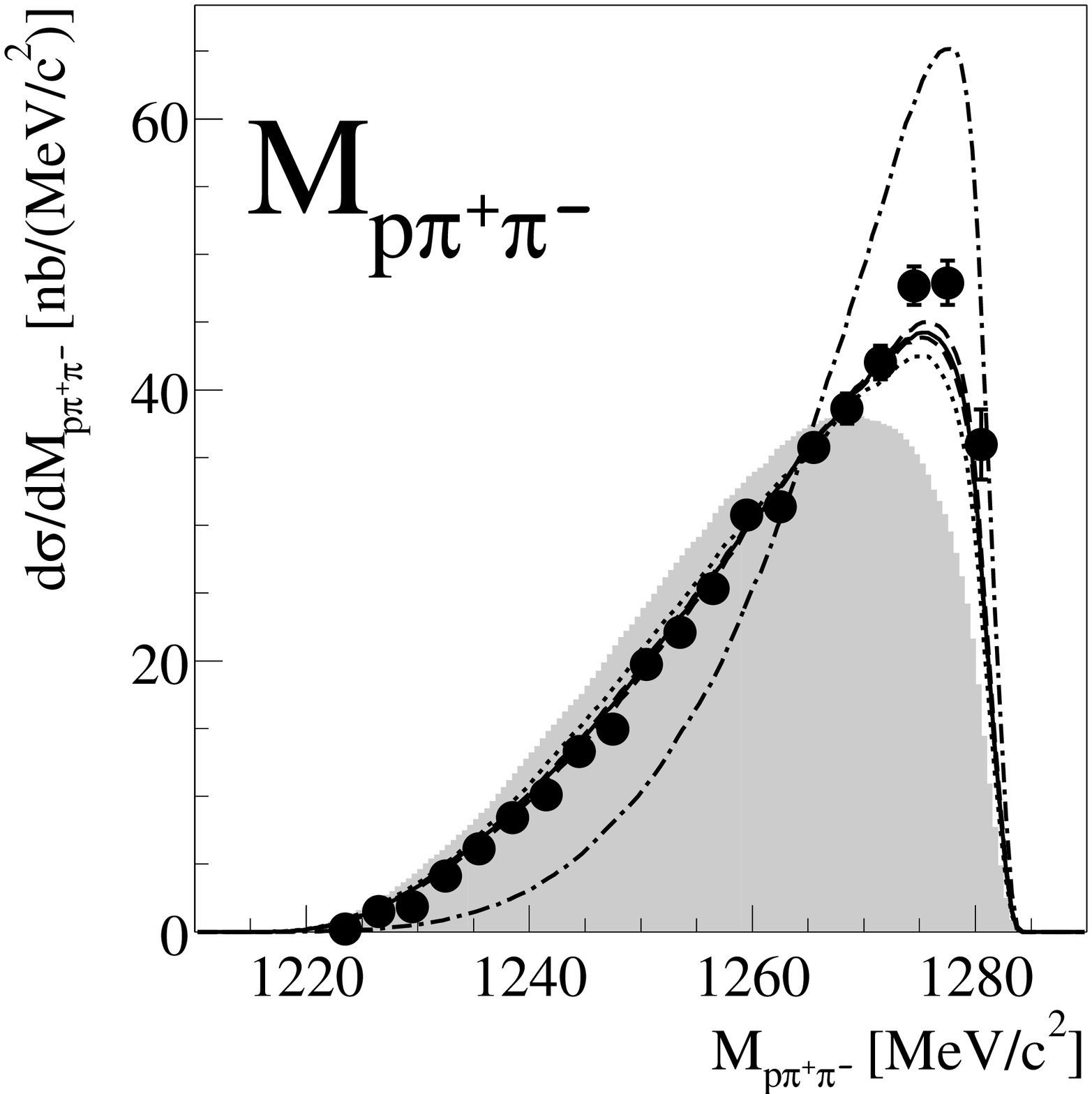}
}
\resizebox{0.2\textwidth}{!}{%
\includegraphics{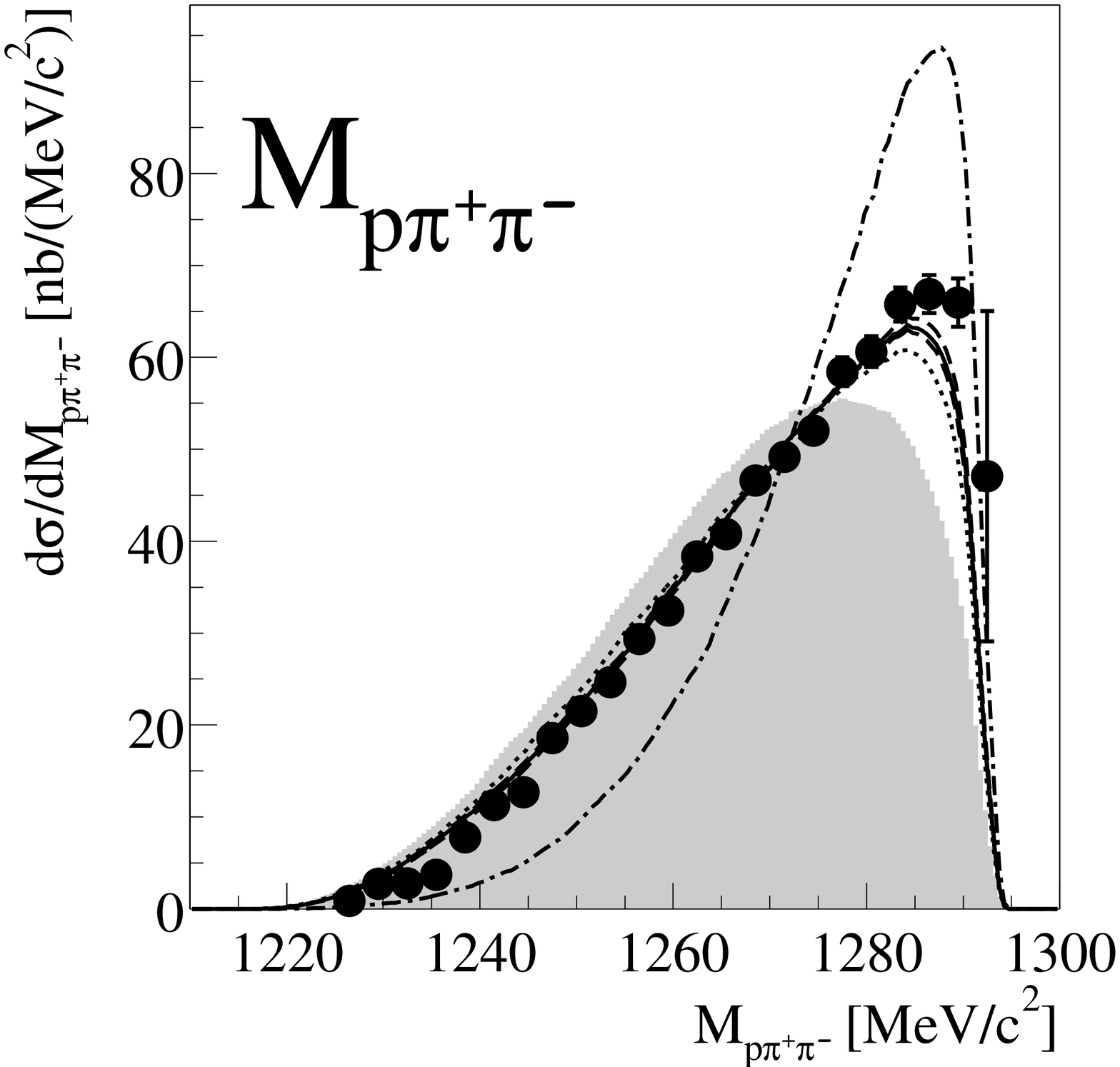}
}

\resizebox{0.2\textwidth}{!}{%
\includegraphics{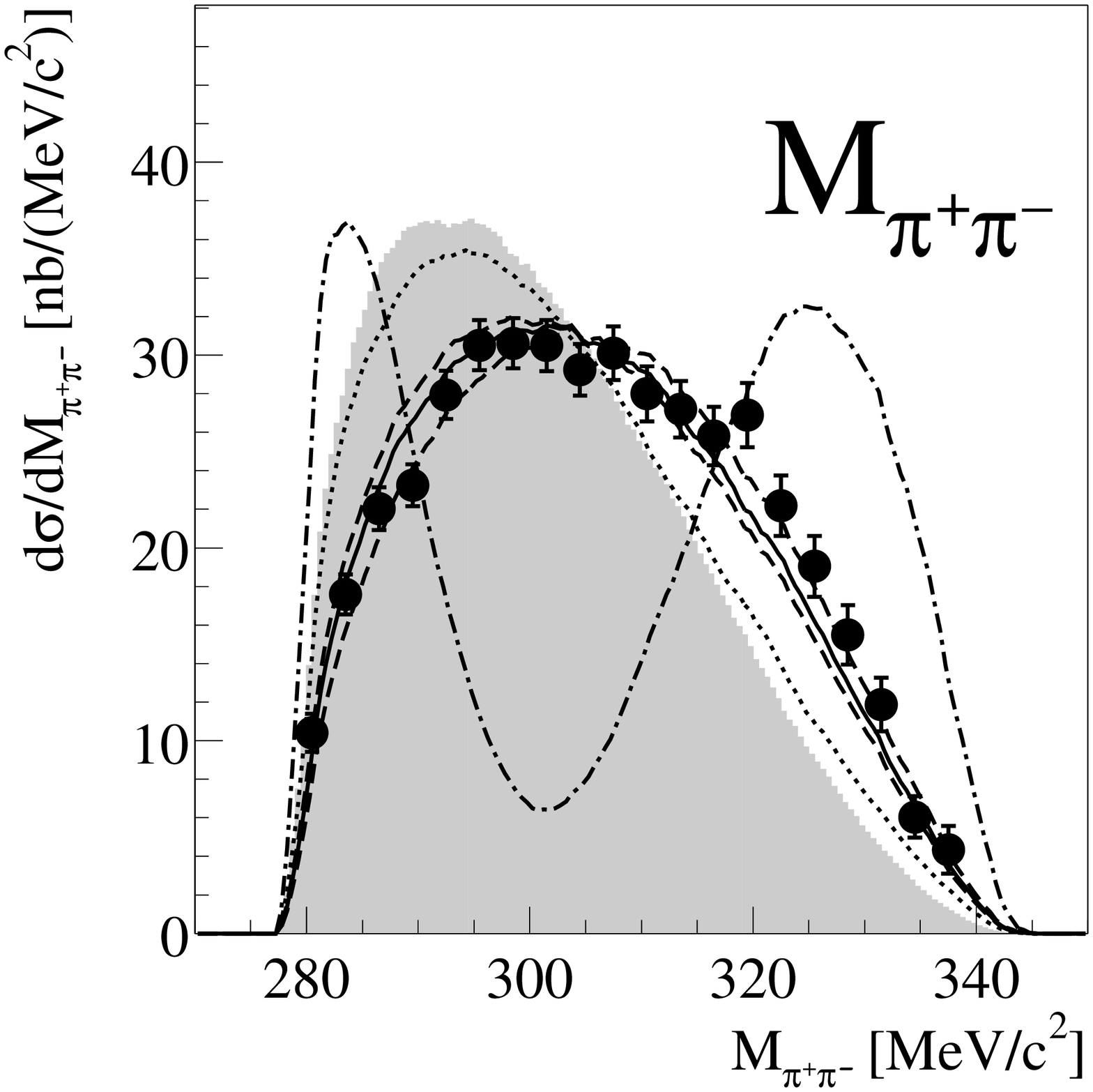}
}
\resizebox{0.2\textwidth}{!}{%
\includegraphics{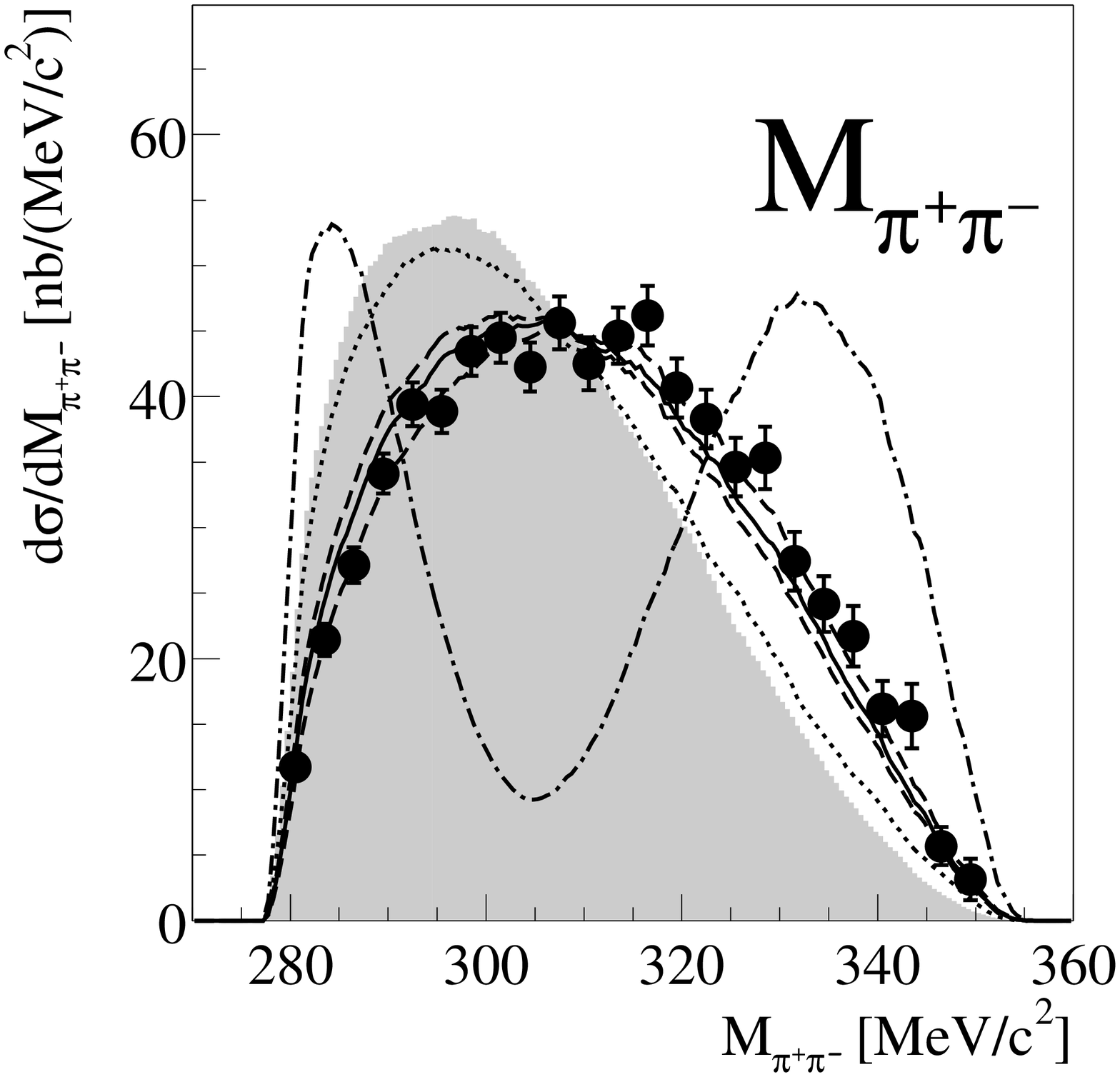}
}

\resizebox{0.2\textwidth}{!}{%
\includegraphics{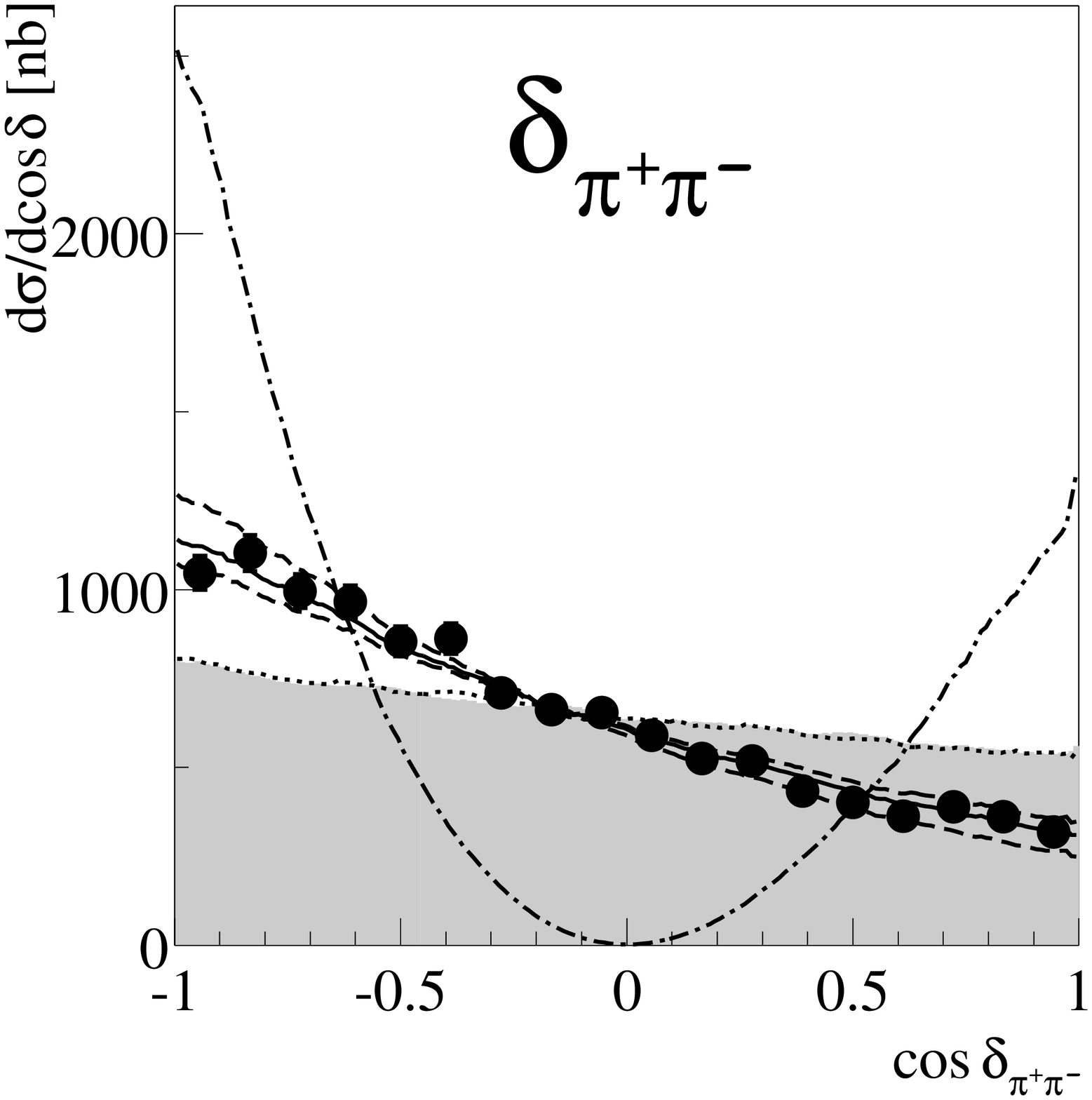}
}
\resizebox{0.2\textwidth}{!}{%
\includegraphics{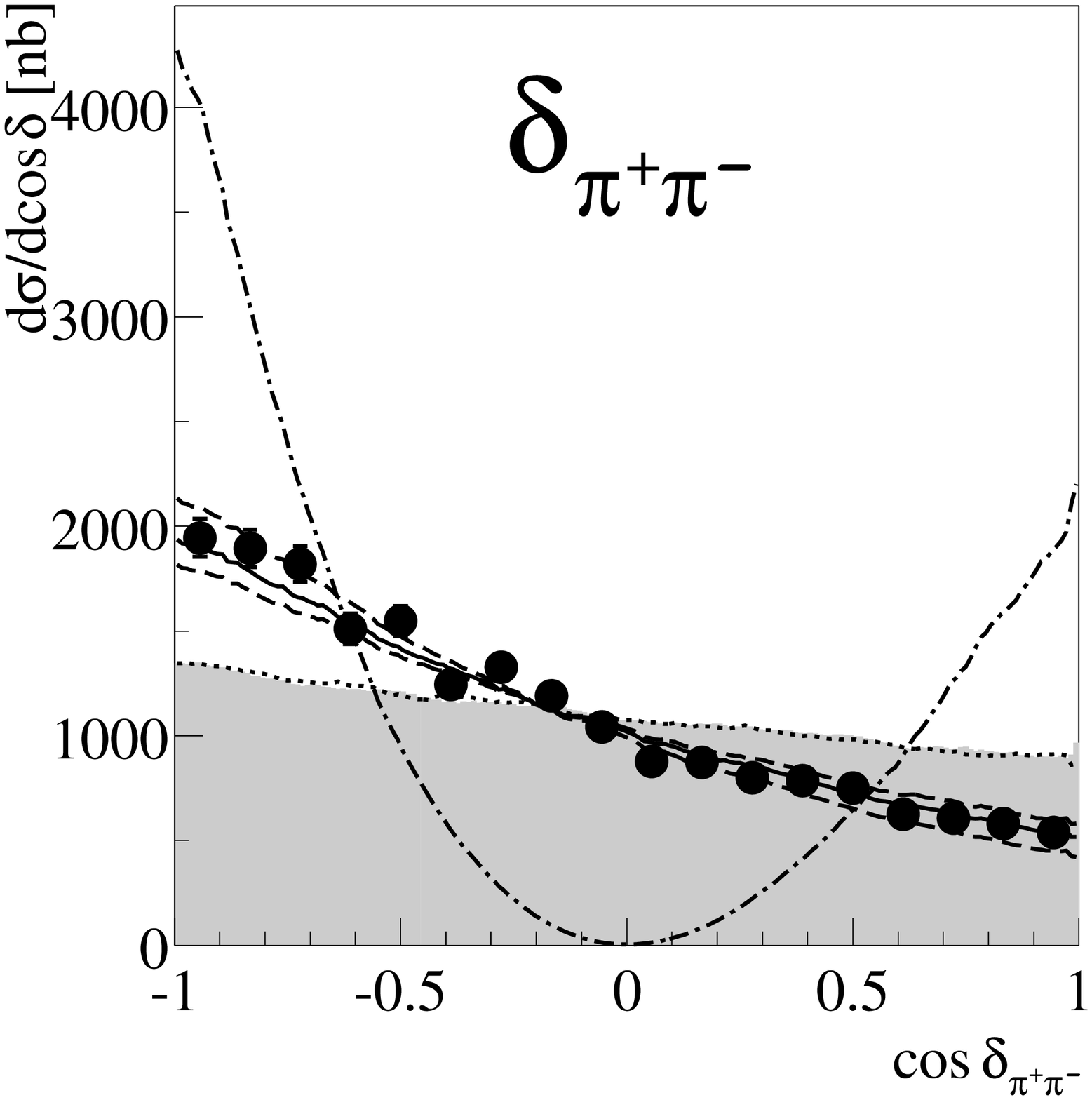}
}
\end{center}
\caption{Influence of the Roper resonance decay onto the differential
cross sections for the invariant masses $M_{p\pi^+\pi^-}$ and $M_{\pi^+\pi^-}$
as well as for the opening angle $\delta_{\pi^+\pi^-}$ between both pions in
the reaction $pp \to pp\pi^+\pi^-$ at $T_p = 750$ MeV (left) and $T_p = 775$
MeV (right). Pure phase space calculations are shown by the shaded area,
dotted lines show the case of a pure $N^\ast \to N\sigma$
decay, whereas the dash-dotted lines exhibit the
scenario for a pure $N^\ast \to \Delta\pi \to N(\pi^+\pi^-)_{I=\ell=0}$
decay. Solid and dashed curves finally show calculations assuming interference
from both decay routes with $a = -0.20, -0.25$ and $-0.33$ \cite{pae02}.
}
\label{fig:1}       
\end{figure}

In order to illustrate the sensitivity of the data to the mixing of both
routes we show calculations for pure phase space, pure transitions $N^\ast \to
N\sigma$ and $N^\ast \to \Delta \pi \to N\sigma$ as well as mixed
scenarios corresponding to $a = -0.20, - 0.25$ and $- 0.33$. The negative sign
reflects the destructive interference between both routes required by the
data.

\section{Results}
\label{sec:3}
From a fit to the data the coefficient $c$ can be
determined and thus also the ratio $R(M_{N\pi\pi})$ of the partial decay
widths for the routes $N^\ast \to \Delta\pi \to N\pi\pi$ and $N^\ast \to
N\sigma$ in dependence of the average $N\pi\pi$ mass populated in the
reaction. For $T_p = 750$ MeV we have $< M_{N\pi\pi} > = 1264$ MeV and for $T_p
= 775$ MeV $< M_{N\pi\pi} > = 1272$~MeV. Having determined $c$ from the data 
we obtain $R(1264) = 0.040(4)$ and $R(1272) =
0.060(6)$. This result just reflects the fact, that in this very low-energetic
tail of the Roper the $N^\ast \to N\sigma$ decay is  by far the dominant
route. However, due to its $k_1 \cdot k_2$ dependence the $N^\ast \to \Delta
\pi$ route is rapidly growing, and
finally will take over at higher energies. If we assume ansatz (1) to be
valid also at higher energies with $c$ being energy independent, then we may
extrapolate to the nominal Breit-Wigner resonance pole. The resulting value,
$R(1440) = 
3.9(3)$, compares very favorably with the PDG value of 4(2). Statistically our
value is much more precise, however, it is still an extrapolation and depends
on the validity of ansatz (1) also at higher energies and hence can not be
considered model-independent. However, as we have demonstrated, the $pp \to
pp\pi^+\pi^-$ reaction offers the possibility to determine experimentally this
ratio at the pole with good precision by measurements at appropriate higher
energies. This reaction, moreover, provides a tool to map out the energy
dependence of the $N^\ast \to N\pi\pi$ decay by successive increase of the
incident proton energy --- a program which is currently pursued at
CELSIUS-WASA and COSY-TOF.

This work has been supported by DFG (Europ. Gra\-du\-ier\-ten\-kol\-leg 683)
and BMBF (06 TU 987).

%
%

\newpage
%

\end{document}